# Assessing Disaster Impacts on Highways Using Social Media: Case Study of Hurricane Harvey

Yudi Chen[1]; Qi Wang, Ph.D.[2]; and Wenying Ji, Ph.D.[3]

[1] Department of Civil, Environmental and Infrastructure Engineering, George Mason University, Fairfax, VA 22030; PH (513) 641-8135; Email: ychen55@gmu.edu

[2] Department of Civil and Environmental Engineering, Northeastern University, Boston, MA 02115; PH (617) 373-7984; Email: q.wang@northeastern.edu

[3] Department of Civil, Environmental and Infrastructure Engineering, George Mason University, Fairfax, VA 22030; PH (703) 993-6130; Email: wji2@gmu.edu

**ABSTRACT**

During and after disasters, highways provide vital routes for emergency services, relief efforts, and evacuation activities. Thus, a timely and reliable assessment of disaster impacts on highways is critical for decision-makers to quickly and effectively perform relief and recovery efforts. Recently, social media has increasingly been used in disaster management for obtaining a rapid, public-centric assessment of disaster impacts due to its near real-time, social and informational characteristics. Although promising, the employment of social media for assessing disaster impacts on highways is still limited due to the inability of extracting accurate highway-related data from social media. To overcome this limitation, a systematic approach is proposed to identify highway-related data from social media for assessing disaster impacts on highways, and a case study of Hurricane Harvey in Houston, TX is employed for the demonstration. The approach is constructed through three steps: (1) building data sources for social media and highways of interest in Houston, respectively; (2) adapting the social media data to each highway through a developed mapping algorithm; (3) assessing disaster impacts through analyzing social media activities in terms of their intensity, geographic, and topic distributions. Results show that the proposed approach is capable of capturing the temporal patterns of disaster impacts on highways. Official news and reports are employed to validate the assessed impacts.

**INTRODUCTION**

Highways play a crucial role in disaster management (e.g., planning evacuation routes (Hobeika and Kim 1998), and distributing relief and recovery resources (Scanlon 2003)) for mitigating or even avoiding disastrous impacts. However, the increasing frequency and intensity of natural disasters (Wallemacq et al. 2015) and poor conditions of infrastructures in the U.S. (ASCE 2017) make highways vulnerable to the massive disruptions caused by natural disasters. Therefore, a timely and reliable assessment of disaster impacts is essential for decision-makers to quickly and effectively plan relief and recovery efforts. Generally, disaster impacts on highways are assessed through physical sensors and field inspections operated by agencies. Although these assessing approaches may be covered with extensive networks, they have limited effects due to the damage to physical sensors and lack of access to field inspections in disasters (Tien et al. 2016).

Recently, the growing adoption of social media platforms, such as Twitter and Facebook, has offered valuable opportunities for consuming public comments and opinions. These social media posts are useful for gaining near real-time, public-centric assessment of the disaster impacts as it unfolds. However, social media data is so abundant that it is necessary to examine through hundreds of thousands, even millions of posts to identify useful information (Imran et al. 2018). To address this issue, supervised classification techniques which require labeled dataset for training are usually used to filter out irrelevant information and classify relevant information into categories of interest (Imran and Castillo 2015). Nevertheless, the performance of these supervised classification techniques is limited due to the scarcity of high-quality training datasets, especially in the first hours of a disaster when there is not enough labeled data for training (Imran et al. 2018). Other than this limitation, the classification categories employed in previous research efforts are focused on general disaster impacts (e.g., affected people, infrastructure damage, and advice) (Imran et al. 2013), which is incapable of providing accurate and comprehensive data for assessing disaster impacts on highways. Therefore, a robust and accurate extraction of highway-related data from social media is essential for obtaining a reliable assessment of disaster impacts on highways.

In this study, a systematic approach is proposed to extract highway-related data from social media to reliably and comprehensively assess disaster impacts on highways. Explicitly, this approach (1) develops a highway-specific mapping algorithm for the effective and accurate extraction of highway-related data from social media; (2) assesses disaster impacts on highways through analyzing the social media activities in terms of their intensity, geographic and topic distributions; and (3) provides a systematic approach for assessing disaster impacts on highways using social media. The content of this paper is organized as follows. In the methodology section, details of building data sources, and data adaption are introduced through the case of Hurricane Harvey in Houston and the social media activities on Twitter. In the results section, the assessed disaster impacts are fully explained, and official reports and news are used to validate the credibility. In the conclusion section, research contributions, limitations, and future work are discussed.

**METHODOLOGY**

The proposed approach is schematically depicted in Figure 1, which mainly consists of three parts: data source, data adaption, and impact assessment. In this manner, the approach pursues to identify highway-related data from social media for obtaining a reliable and comprehensive assessment of disaster impacts on highways. The approach is examined using the case study of Hurricane Harvey which inflicted $125 billion in damage, primarily from catastrophic rainfall-triggered flooding in the Houston metropolitan area and Southeast Texas (Blake and Zelinsky 2018). As the principal city in the Houston metropolitan area, Houston is selected as the area of interest for assessing disaster impacts on highways because of its potential for a strong social media activity. Additionally, the Twitter platform is employed in this

study for providing social media data due to its easy access for collecting large-scale dataset.

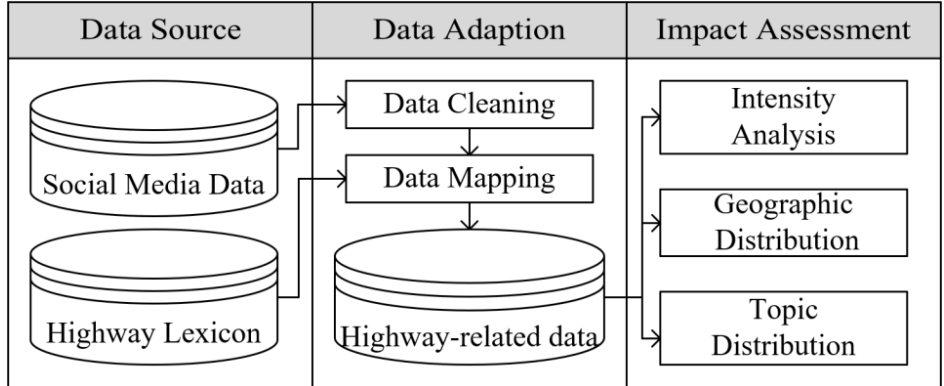

Figure 1. The research approach for assessing disaster impacts on highways

**Data Source**

This study primarily involves two types of data: social media data (collected from Twitter) and highway lexicon (built with the information of highways in Houston).

*Social Media Data*

The Twitter streaming API is used for collecting geotagged tweets (Wang and Taylor 2015). Two types of filters, time period and location bounding box, are employed to further filter the collected geotagged tweets to ensure that the tweets used in this study are posted in the time of Hurricane Harvey and located in Houston. The time period is two weeks from Aug. 23, 2017 (three days before the start of the unprecedented rainfall brought by Harvey) to Sep. 5, 2017 (one week after the end of the rainfall). The location bounding box is comprised of latitudes and longitudes of boundaries for the area of interest. In this case, the bounding box is created to cover the urban areas of Houston, with corresponding latitudes [29.427926, 30.157266] and longitudes [-95.902705, -94.997805]. Finally, 53,567 tweets were filtered for extracting highway-related data.

Harvey brought extensive rainfall in Houston where many locations observed record amounts of daily rainfall (Blake and Zelinsky 2018). Figure 2 shows the number of daily tweets (represented by blue squares) and the amount of daily rainfall in Houston (represented by orange circles) obtained from the National Oceanic and Atmospheric Administration (NOAA). Similar trends are observed in the number of daily tweets and the amount of daily rainfall. The number of daily tweets increased significantly from Aug. 26 to Aug. 27 due to the record rainfall brought by Harvey. Unlike the sharp decline of rainfall, the daily tweets remained at a high level of activity from Aug. 27 to Aug. 30 due to the rainfall-triggered flooding. Given these temporal variations, three phases (i.e., pre-peak, peak and post-peak) are defined to capture the temporal patterns of disaster impacts on highways. Explicitly, the period from Aug.23 to Aug. 25 is defined as the pre-peak phase since there was neither rainfall nor the strong activity of daily tweets; period of Aug. 26 to Aug. 30 is defined as the peak phase due to the extensive rainfall or the strong activity of daily tweets;

and period of Aug. 31 to Sep. 5 is defined as the post-peak phase as the activity of daily tweets decreased to a low level.

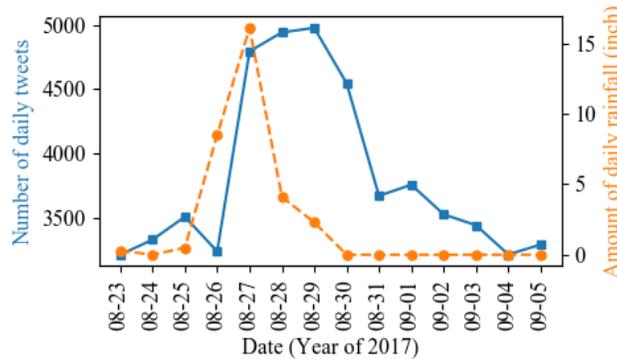

**Figure 2. Number of daily tweets and rainfall in Houston**

*Highway Lexicon*

Highways of interest in Houston are determined by their annual average daily traffic (AADT) and locations in Houston. Two beltways (I-610 and Sam Houston Tollway (SHT)) and three interstate highways (I-10, I-45, and I-69) are selected as they go through the urban areas of Houston, with considerable AADT. Geographic map of the highways is shown in Figure 3.

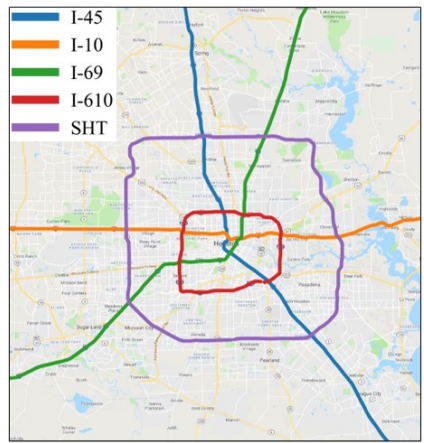

**Figure 3. Geographic map of the selected highways**

Since tweets are usually brief, informal, and often contain slang, grammar mistakes and abbreviations, the semantic analysis conducted through natural language processing is limited for identifying useful information (Congosto et al. 2015). Therefore, in this study, manual exploration of the key search terms for identifying highway-related data from social media is performed with the incorporation of tweets' characteristics (e.g., abbreviations) and highway nicknames (e.g., beltway 8 for Sam Houston Tollway). Two types of search terms are finally defined, thus direct search terms and indirect search terms, as shown in Table 1. The direct search terms actually are official designations and nicknames of highways, while the indirect search terms are composed by a highway-specific numeric value (e.g., 45, 69 and 10) and a location term (e.g., gulf, north, and katy). Meanwhile, the abbreviations of the location terms are included as they are frequently used on Twitter. Detailed employment of these terms for identifying highway-related data is introduced in the data mapping section.

Table 1. Search terms

| Highway | Direct Search Terms | Indirect Search Terms |
|---|---|---|
| I-45 | "i-45", "i45" | "45", "45 gulf", "45 north", "45 n" |
| I-10 | "i-10", "i10" | "10", "10 baytown east", "10 baytown e", "10 katy" |
| I-69 | "i-69", "i69" | "69", "69 eastex", "69 southwest", "69 sw" |
| I-610 | "i-610", "i610" | "610", "610 west", "610 east", "610 north", "610 south", "610 w", "610 e", "610 n", "610 s" |
| SHT | "beltway 8", "beltway8", "belt8" | "sam houston" |

**Data Adaption**

Adapting the collected raw data into readable data is essential in data mining as the raw data are usually unstructured and often too noisy to be utilized for decision support. In this section, the collected tweets are adapted to each highway through two processes: data cleaning and data mapping.

*Data Cleaning*

Data cleaning, a process of detection and removing errors and inconsistencies, is required for mining social media data as it usually contains excessive amounts of slang and emoticons (Zhu and Cai 2015). The cleaning pipeline in this study is constructed in six steps:

Step 1. Remove useless URL links since this study is based on text information;

Step 2. Perform tokenization to split a tweet into separate terms based on a set of delimiters;

Step 3. Transform all terms into their lower case;

Step 4. Perform lemmatization to convert the terms into their root terms, such as "closed" to "close", since they express the same information of disaster impacts;

Step 5. Remove useless and invalid symbols through regular expressions;

Step 6. Remove stopping terms that are most common terms (e.g., is, of and often) for enhancing the representativeness of remaining terms.

With these cleaning steps, each tweet is represented by a set of enhanced terms. This enhanced representation for each tweet ensures the efficiency and accuracy of the mapping algorithm developed in the following section.

*Data Mapping*

To assess disaster impacts on each highway, the cleaned tweets are mapped accordingly through a highway-specific mapping algorithm. In the algorithm, the two types of search terms (i.e., direct and indirect terms) of each highway are compared with a cleaned tweet to determine whether the tweet is related to this specific highway. For direct search terms, if a tweet contains one of the direct search terms for a specific highway, then this tweet is related to this specific highway. While for indirect search terms, a tweet contains one of the indirect search terms only indicates that there is a high potential of relatedness, but further demonstrations are still needed. For example, in the tweet "It's like 45 songs that isn't R&B RT @Jdxthompson: 2000's R&B was the best", although it contains "45" which is one of the indirect search terms for I-45, it is not related to I-45. A set of highway terms ("highway", "hwy", "freeway", "fwy", "tollway", "tlwy", "parkway", "pwy", "loop", "lp") which are exclusively used for highways is defined for further demonstration. The algorithm is illustrated in a step-by-step form and repeats for each tweet.

Given a tweet, name it as *the_tweet*:
Step 1. Select a highway of interest and name it as *the_highway*;
Step 2. Load the corresponding direct and indirect search terms and name them as *direct_terms* and *indirect_terms* respectively;
Step 3. Check if there is an intersection between *the_tweet* and *direct_terms*. If yes, *the_tweet* is related to *the_highway*, and go to Step 7;
Step 4. Check if there is an intersection between *the_tweet* and *indirect_terms*. If yes, name the intersection as *indirect_intersection*, otherwise, go to Step 7;
Step 5. Find the adjacent terms of *indirect_intersection* from *the_tweet*, and name it as *neighbors*;
Step 6. Check if there is an intersection between *neighbors* and *highway_terms*. If yes, *the_tweet* is related to *the_highway*;
Step 7. Go to Step 1 and repeat until all highways of interest are selected.

With this highway-specific mapping algorithm, the tweets are mapped accordingly to assess disaster impacts on each of the highways based on their respective social media activities.

**RESULTS**

In this section, disaster impacts on the highways are assessed through analyzing their social media activities from perspectives of intensity, geographic and topic distributions. By doing so, a reliable and comprehensive assessment of disaster impacts on highways is obtained.

**Intensity Analysis**

The severity of disaster impacts on highways is assessed in a comparative and qualitative manner through the intensity analysis. To minimize the bias of highway situations in normal cases, the intensity is calculated by normalizing the average number of daily tweets with the average number in the pre-peak phase where Harvey had not yet impacted Houston. The intensities of the highways are illustrated in Figure 4. The intensities are the same for all highways in the pre-peak phase due to the normalization, while in the peak and the post-peak phases, intensity differences between the highways were obvious as Harvey affected them differently. A stronger intensity indicates more severe disaster impacts and vice versa. For example, I-10 and SHT had stronger intensities than other highways in the peak and the post-peak phases, which indicates that these two highways had more severe impacts and required a longer time for recovery. This is validated by the news of road closures posted by Houston Chronicle, the I-10 and SHT were two highways with more closures and high water locations than the other three highways (Barron and Hill 2017).

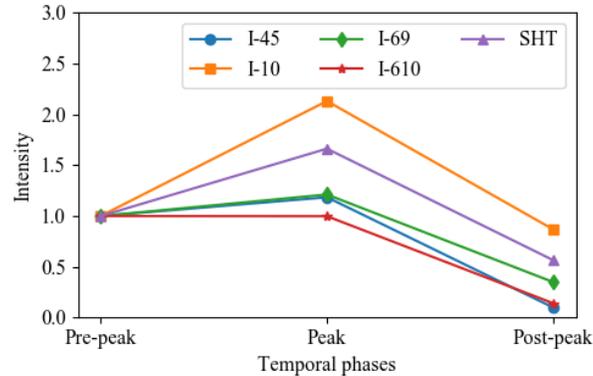

**Figure 4. The intensities of social media activities**

**Geographic Distribution**

Similar with the intensity analysis, the geographic patterns of disaster impacts are obtained through comparing the distributions in the peak and post-peak phases with the distribution in the pre-peak phase to minimize the bias of highway situations in normal cases. Figure 5 shows the geographic distributions of the social media activities for each highway in the three temporal phases. Obviously, the distribution in the peak phase is more widely spread than the distribution in the pre-peak phase, which indicates that the disaster impacts on highways were geographically widespread. Take I-10 as an example, tweets were mainly located around some specific locations (e.g., intersection) in the pre-peak phase, while in the peak phase, tweets were distributed across all sections of I-10. Another phenomenon to notice is that the westside of the highways had more tweets than the eastside in the post-peak phase, especially for I-10. This is consistent with the facts that the westside of Houston required a long-time recovery as it was severely flooded by the water release from two flood-control reservoirs, Barker and Addicks, which are located in the westside of Houston (Lindner and Fitzgerald 2018).

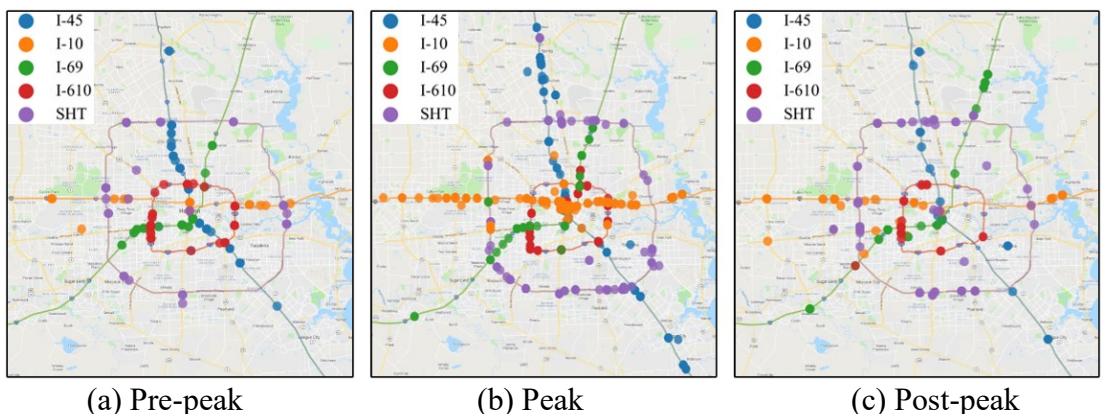

(a) Pre-peak               (b) Peak               (c) Post-peak
**Figure 5. Geographic distributions of the social media activities**

In addition, almost all highway-related tweets are located on or near their respective highway. This consistency indicates an extremely strong relatedness between the tweets and their related highway, and in turn, it demonstrates the satisfying performance of the developed highway-specific mapping algorithm.

**Topic Distribution**

The analysis of topic distribution is to assess disaster impacts from the contents of the social media activities. The topics of each highway are obtained by extracting the top terms from their own social media activities. The search terms (i.e., direct and indirect) are removed as they are already known information for their related highway. Table 2 lists the top 5 terms for each highway with a frequency in the following bracket. It is noticed that the frequencies of these top terms are fairly large, around half of the number of tweets, for all highways in all temporal phases. This indicates that most tweets are related to these top terms, and therefore the extracted top 5 terms are representative for these highway-related tweets.

Table 2. Topic distributions of the social media activities

| Hwy | Pre-peak | | Peak | | Post-peak | |
|---|---|---|---|---|---|---|
| | # of tweets | Topics | # of tweets | Topics | # of tweets | Topics |
| I-10 | 38 | lane (26) stop (20) accident (19) min (16) delay (16) | 171 | flood(101) lane (99) water (93) high (93) baytown (87) | 186 | close (92) flood (90) katy (80) westside (61) frontage (52) |
| I-45 | 65 | stop (54) back (47) min (46) delay (46) accident (45) | 176 | water (117) high (113) lane (94) close (85) flood (79) | 66 | lane(8) accident (7) flood (7) downtown (7) close (15) |
| I-610 | 53 | stop (42) back (37) delay (34) accident (34) min (34) | 138 | water (79) high (79) lane (69) affect (66) flood (63) | 39 | accident (24) stop (21) back (18) block (15) delay (13) |
| I-69 | 30 | stop (25) accident (23) back (21) delay (18) min (18) | 95 | flood (54) close (51) water (39) inbound (38) high (37) | 48 | stop (22) accident (18) back (17) inbound (16) close (16) |
| SHT | 42 | stop (23) lane (22) back (21) block (17) accident (17) | 150 | flood (122) close (82) exit (56) high (52) water (52) | 139 | close (57) westside (50) flood (48) katy (45) frontage (32) |

Disaster impacts on each highway are captured through the temporal evolutions of these topics. The topics of all highways were dominated by two types of topics, daily-traffic terms (e.g., "stop", "accident", and "delay") and Harvey-impact terms (e.g., "flood", "water", and "close"). In the pre-peak phase, topics were dominated by the daily-traffic terms as Harvey had not yet affected Houston, while in the peak phase, the dominated topics turned to Harvey-impact terms as highways were flooded or with high water. In the post-peak phase, topics for I-10 and SHT were still dominated by the Harvey-impact terms, which indicates a slow process of recovery, while topics for I-610 and I-69 turned back to the daily-traffic terms, which suggests a rapid recovery. Unlike other highways in which topics were dominated by only one type, both the daily-traffic terms (e.g., accident) and the Harvey-impact terms (e.g., flood) existed in I-45. This combined involvement reflects a recovery process in which I-45 was flooded in the beginning and recovered at the end of the post-peak phase.

Besides these two dominated terms, location terms in the topics are capable of pointing out the areas that are severely affected. For instance, "katy" and "westside" in the topics for I-10 and SHT in the post-peak phase indicate that the west sides of these two highways are severely affected, which is demonstrated in the analysis of geographic distribution.

**CONCLUSION**

Social media, with its near real-time, social and informational characteristics, is playing an increasingly important role in disaster management. In this study, we propose a systematic approach for extracting accurate highway-related data from social media to assess disaster impacts on highways. The case of Hurricane Harvey in Houston and the social activities on Twitter are employed for the demonstration. Results show that the proposed approach is capable of extracting highway-related data accurately and assessing the disaster impacts on highways reliably and comprehensively.

This study contributes to academia by (1) developing an effective and reliable mapping algorithm for identifying highway-related data from social media; (2) assessing disaster impacts on highways through a comprehensive analysis of social media activities; and (3) proposing a systematic approach for pipelining the assessment of disaster impacts on highways using social media. For practitioners, the assessed disaster impacts can provide a rapid and reliable awareness of highway situations for effective planning of relief and recovery efforts.

Although this study proposed a systematic approach to identify highway-related data for assessing disaster impacts, it is unable to classify the data based on its impacts on highways, such as "block" and "close" are way different for highway traffic. Therefore, classifying these impacts to a set of detailed categories using machine learning algorithms will be studied in further research.

**ACKNOWLEDGMENT**

This study is supported by the Thomas F. and Kate Miller Jeffress Memorial Trust and the National Science Foundation (Grant No. 1761950). Any opinions, findings, and conclusions or recommendations expressed in this material are those of the authors and do not necessarily reflect the views of the Thomas F. and Kate Miller Jeffress Memorial Trust and the National Science Foundation. The authors also gratefully acknowledge the support of internal grants from the Global Resilience Institute, Northeastern University.